\documentclass[twocolumn,showpacs,aps,prl,floatfix,%
superscriptaddress]{revtex4}  

\usepackage{graphicx} 

\newcommand{\ee}{\end{equation}} 
\newcommand{\eq}{{\,=\,}} 
\newcommand{\bm}{\bf}

\begin{document} 
 
 
\title{Conical flow due to partonic jets in central Au+Au collisions}
 
\date{\today}
  
\author{A. K. Chaudhuri} 
 
\email{akc@veccal.ernet.in} 
\affiliation{Variable Energy Cyclotron Centre, 1-AF, Bidhan Nagar,  
Kolkata - 700 064, India} 
 
\begin{abstract} 

In jet quenching, a hard QCD parton, before fragmenting into a jet of hadrons, deposits a fraction of its energy in the medium, leading to  suppressed production of high-$p_T$ hadrons. The
process can generate shock waves. 
We study the distortion of Mach shock waves due to jet quenching in central Au+Au collisions and its effect on particle production.
Finite fluid velocity and inhomogeneity of the medium can 
distort the Mach shock front significantly such that
the inside shock front disappear and the outside shock front
is opened up.  
We also show that the STAR data on azimuthal distribution of
background subtracted secondaries, associated with high
$p_T$ trigger, are reasonably well explained by the excess  pions
produced due to partonic energy loss.

\end{abstract} 
 
 
\maketitle 
 

The three most important results  that came out from the heavy ion
programme at RHIC are  (i) dramatic suppression of inclusive hadrons production at large transverse momentum (high $p_T$ suppression) \cite{Adcox:2001jp,Adler:2002xw,Adams:2003kv},
(ii)disappearance of away side two hadron correlation peak
\cite{STARjetqu} and   (iii) large elliptic flow \cite{elliptic}.  
High $p_T$ suppression confirmed the theoretical prediction of jet quenching \cite{QGP3jetqu} . 
Long before the RHIC Au+Au collisions, it was predicted, in a pQCD calculation, that in a deconfined medium, high-speed partons will suffer energy loss, leading to suppressed production of hadrons. The observed high $p_T$ suppression in Au+Au collisions are in agreement with the prediction.  Large elliptic flow observed in non-central Au+Au collisions confirms fluid like behavior of the produced matter. The elliptic flow data  are well explained in an
ideal hydrodynamic model, with initial energy density of deconfined matter $\varepsilon_i \sim$ 30 $GeV/fm^{3}$, thermalized at an initial time $\tau_i$=0.6 fm \cite{QGP3v2}. 
Transport based models, on the other hand,  require unrealistically large cross-section to reproduce the observed elliptic flow \cite{Molnar:2001ux}.  
All these observations are being treated 
as evidences for the creation of a very dense, color opaque medium of deconfined quarks and gluons \cite{QGP3jetqu}.

As the jet quenching models suggest, if the partons lose energy in the medium, a natural question arises, what happens to the
lost energy? In \cite{shuryak}, it has been suggested that
a fraction of the lost energy goes to collective excitation, called
"conical flow".   The quenching jet moves at
speed of light ($c_{jet}\approx 1$), much larger than the speed of sound of the medium ($c^2_s < 1$). The quenching jet can produce a shock wave with 
Mach cone angle, $\theta_M=cos^{-1} c_s/c_{jet}$. The resulting conical flow will
have characteristic peaks at $\phi=\pi-\theta_M$ and $\phi=\pi+\theta_M$. Indications of such peaks are seen in azimuthal
distribution of secondaries associated with high $p_T$ trigger in
central Au+Au collisions \cite{Wang:2004kf,Jacak:2005af}. As Mach cone is sensitive to the speed
of sound of the medium, it raises the possibility of measuring
the speed of sound of the deconfined matter of Quark-Gluon-Plasma.
 
Recently, we have (numerically) solved  hydrodynamical equations with an additional source, representing the quenching jet \cite{Chaudhuri:2005vc}. We concluded that for realistic jet energy loss, the detailed structure seen in the azimuthal correlation of secondaries, can not be explained 
by Mach cone formation. However, that simulation was limited to b=0 collision. More importantly, only one trajectory (along the diagonal) was considered. Recently, distortion of Mach regions,
due to radial and longitudinal expansion was studied \cite{Satarov:2005mv}. Finite fluid velocity induces significant deformation of Mach shocks.  
For the perturbation (jets) moving along the diagonal, the Mach cone
 opening angle ($\tilde{\theta}_M=sin^{-1}c_s/c_{jet}=\pi/2-\theta_M)$
is reduced, but it remains symmetric,
$\tilde{\theta}_{M+} \eq \tilde{\theta}_{M-}$,  where, $\tilde{\theta}_{M+}$ and $\tilde{\theta}_{M-}$ are the upper and lower cone opening angles, with respect to the jet.
Mach cone become asymmetric when the jet moves along a chord ($\tilde{\theta}_{M+} \neq \tilde{\theta}_{M-}$). The relation between
$\tilde{\theta}_{M+}$ and $\tilde{\theta}_{M-}$ reversed as the jet moves from upper half plane to lower half plane.
 It is important see,
whether exact numerical simulation corroborate such 
predictions. Such changes in Mach angle, will also affect the particle production as opposed to  the only diagonal jets studied in \cite{Chaudhuri:2005vc}.

\begin{figure}[h] 
\includegraphics[bb=14 13 581 829,width=0.80\linewidth,clip]{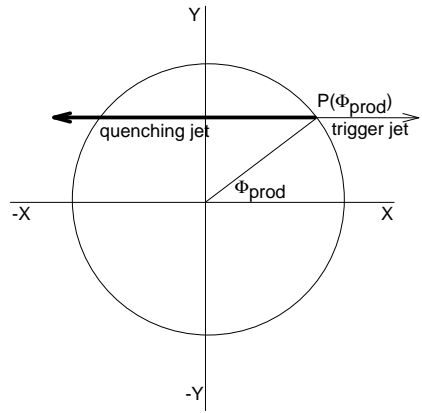}
\vspace{-4.5cm}
\caption{Schematic representation of a jet moving through the
medium. The high $p_T$ pair is assumed to produce on the surface of the fireball characterized by the angle
$\phi_{prod}$. One of the jet escapes forming the trigger jet, the
other enters into the fireball.}
\label{F1}
\end{figure}

A schematic representation
of the jet moving through the medium is shown in Fig.\ref{F1}.
We assume that just before hydrodynamics become applicable, a pair 
of high-$p_T$ partons is produced at $P$.  
One of jet moves outward  and escapes, forming the trigger jet. The other enters into the fireball. For jet trajectory parallel
to the x-axis, the production point can be characterised by
the angle $\phi_{prod}$ varying between [$-\pi/2,+\pi/2$].
Strong jet quenching and survival of the trigger jet, forbids production
in the interior of the fireball. Jet pairs can be produced only on a thin shell on the surface of the fireball. For an impact parameter $b$ collision, we assume that the pair is produced on the surface of the ellipsoidal fireball with minor and major
axis, $A=R-b/2$ and $B=R \sqrt{1-b^2/4R^2}$ with
R=6.4 fm (for Au+Au collisions).  
The fireball is expanding and cooling. The ingoing parton travels 
at the speed of light and loses energy in the fireball which 
thermalizes and acts as a source of energy and momentum for 
the fireball medium.

The model we use
is described in \cite{Chaudhuri:2005vc}. Briefly,  
 we solve the energy-momentum conservation equation,

\begin{equation} \label{1}
\partial_\mu T^{\mu\nu}=J^\nu,
\end{equation}

\noindent where the source is modeled as,

\begin{eqnarray} 
\label{2} 
&&J^\nu(x)=J(x)\,\bigl(1,-1,0,0\bigr),\\
\label{3}
 &&J(x) = \frac{dE}{dx}(x)\, \left|\frac{dx_{\rm jet}}{dt}\right| 
          \delta^3(\bm{r}-\bm{r}_{\rm jet}(t)).
\end{eqnarray} 
Massless partons have light-like 4-momentum, so the current $J^\nu$
describing the 4-momentum lost and deposited in the medium by the 
fast parton is taken to be light-like, too. $\bm{r}_{\rm jet}(t)$ is 
the trajectory of the jet moving with speed $|dx_{\rm jet}/dt|\eq{c}$.
$\frac{dE}{dx}(x)$ is the energy loss rate of the parton as it moves 
through the liquid. It depends on the fluid's local rest 
frame particle density. Taking guidance from the 
phenomenological analysis of parton energy loss observed in Au+Au 
collisions at RHIC \cite{Eloss} we take
\begin{equation}
\label{4}
  \frac{dE}{dx} = \frac{s(x)}{s_0} \left.\frac{dE}{dx}\right|_0
\end{equation}
where $s(x)$ is the local entropy density without the jet. 
The measured suppression of high-$p_T$ particle production in Au+Au 
collisions at RHIC was shown to be consistent with a parton energy 
loss of $\left.\frac{dE}{dx}\right|_0\eq14$\,GeV/fm at a reference 
entropy density of $s_0\eq140$\,fm$^{-3}$ \cite{Eloss}. We will
use this realistic jet energy loss through out the present calculation.

For the hydrodynamic evolution we use a modified version of the publicly available hydrodynamic code AZHYDRO \cite{QGP3v2,AZHYDRO}. The code
  is formulated in 
$(\tau,x,y,\eta)$ coordinates, where $\tau{=}\sqrt{t^2{-}z^2}$ is the
longitudinal proper time, 
$\eta{=}\frac{1}{2}\ln\left[\frac{t{+}z}{t{-}z}\right]$ 
is space-time
rapidity, and $\bm{r}_\perp{\,=\,}(x,y)$ defines the plane transverse to the 
beam direction $z$. AZHYDRO employs longitudinal boost invariance 
along $z$ but this is violated by the source term (\ref{3}). We 
therefore modify the latter by replacing the $\delta$-function 
in (\ref{3}) by

\begin{eqnarray}
\label{5}
  \delta^3(\bm{r}-\bm{r}_{\rm jet}(t)) &\longrightarrow&
  \frac{1}{\tau}\,\delta(x-x_{\rm jet}(\tau))\,\delta(y-y_{\rm jet}(\tau))
\nonumber\\
 &\longrightarrow&\frac{1}{\tau} \, 
\frac{e^{-(\bm{r}_\perp-\bm{r}_{\perp,{\rm jet}}(\tau))^2/(2\sigma^2)}}
      {2\pi\sigma^2}
\end{eqnarray}
with $\sigma{\,=\,}0.70$\,fm. 

Intuitively, this replaces the ``needle'' 
(jet) pushing through the medium at one point by a ``knife'' cutting the
medium along its entire length along the beam direction. 
Thus assumption of boost-invariance will over estimate the
effect of jet quenching. 
While a complete study of this would require a full 
(3+1)-dimensional hydrodynamic calculation, the present boost-invariant
simulation should give a robust upper limit for the effect produced by the quenching jet. 

The modified hydrodynamic equations in $(\tau,x,y,\eta)$ coordinates
read \cite{Chaudhuri:2005vc,AZHYDRO}
%
\begin{eqnarray} 
\label{6} 
  \partial_\tau \tilde{T}^{\tau \tau} + 
  \partial_x(\tilde{v}_x \tilde{T}^{\tau \tau}) +
  \partial_y(\tilde{v}_y \tilde{T}^{\tau \tau}) 
  &=& - p + \tilde{J},
\\ 
\label{7} 
  \partial_\tau \tilde{T}^{\tau x} +
  \partial_x(v_x \tilde{T}^{\tau x}) +
  \partial_y(v_y \tilde{T}^{\tau x}) 
  &=& - \partial_x \tilde{p} - \tilde{J}, \quad
\\ 
\label{8} 
  \partial_\tau \tilde{T}^{\tau y} +
  \partial_x(v_x \tilde{T}^{\tau y}) +
  \partial_y(v_y \tilde{T}^{\tau y}) 
  &=& -\partial_y \tilde{p},  \quad
\end{eqnarray}  
%
where $\tilde{T}^{\mu\nu}\eq\tau T^{\mu\nu}$, 
$\tilde{v}_i{\eq}T^{\tau i}/T^{\tau\tau}$,
$\tilde p\eq\tau p$, and $\tilde{J}\eq\tau J$.

We solve these equations for Au+Au collisions at impact 
parameter $b=3$ fm. Roughly it corresponds to 0-5\% centrality
Au+Au collisions. 
We use the standard 
initialization described in \cite{QGP3v2} and provided in the 
downloaded AZHYDRO input file \cite{AZHYDRO}, corresponding to a 
peak initial energy density of $\varepsilon_0\eq30$\,GeV/fm$^3$ at 
$\tau_0\eq0.6$\,fm/$c$. We use the equation of state EOS-Q described 
in \cite{QGP3v2,AZHYDRO} incorporating a first order phase transition 
and hadronic chemical freeze-out at a critical temperature 
$T_c{\,=\,}164$\,MeV. The hadronic sector of EOS-Q is soft with a 
squared speed of sound $c_s^2 \approx 0.15$. 

\begin{figure}[h] 
\includegraphics[bb=14 13 581 829,width=1.0\linewidth,clip]{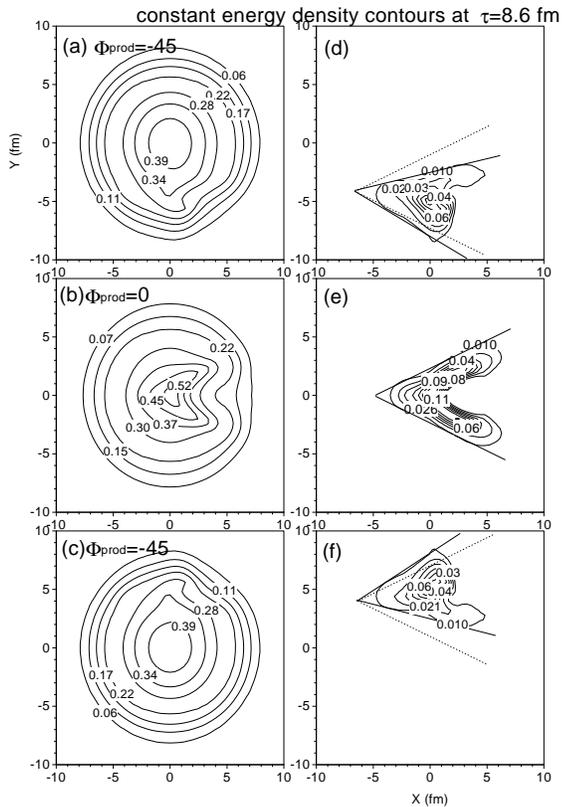}
\vspace{-2cm}
\caption{Left panels (a), (b) and(c) shows the constant energy density contours at time $\tau$=8.6 fm, The jet is produced at
$\phi_{prod}=-45^o$, $0^o$ and $45^o$. The right panels,
(d), (e) and (f) show the excess
energy density due to quenching jet only. The conical structure is clearly visible. Approximate Mach cones are depicted by the
solid lines. To show the distortion, for the off diagonal jets, we have shown the symmetric Mach cone by the dotted lines.}
\label{F2}
\end{figure}

In Fig.\ref{F2}, we have shown the results of our simulations.
We have considered all positions of the jet trajectories,
$-\pi/2 \leq \phi_{prod} \leq \pi/2$. Some representative cases,
jets in the upper half plane ($\phi_{prod}=45^o$), in lower
half plane  ($\phi_{prod}=-45^o$) and along the diagonal, 
are shown in Fig.\ref{F2}.
In the left panels  we have shown the constant energy density contours after $\tau$=8.6 fm of evolution. 
With
realistic energy loss, the perturbation to energy density is small
and conical flow developed by the Mach shock wave is  
barely visible. To see the Mach cone formation, we subtract the
energy density of the fluid evolved under a similar condition, but with out the quenching jet. Contours of  excess energy density, due to the quenching jet, are shown in the right panels, d, e and f. Mach cone like structure is clearly visible. We also note that
the distortion of the Mach cone due to transverse velocity
roughly corresponds to the theoretical study \cite{Satarov:2005mv} . In the figure,
we have drawn the (approximate) Mach cone. For the jet
moving along the diagonal, the Mach region is symmetric
with respect to the jet axis, the upper and lower cone opening angles are same, $\tilde{\theta}_{M+}=\tilde{\theta}_{M-}$. But for the jets moving along
a chord, it is asymmetrical. Due to transverse expansion Mach region is pushed out.  
The inner surface is pushed more than the
outer surface. In the upper half plane, $\tilde{\theta}_{M+} < \tilde{\theta}_{M-}$,
the relation is reversed in the lower plane, $\tilde{\theta}_{M+} > \tilde{\theta}_{M-}$. While the simulations corroborate the 
predicted distortions of Mach shock regions, we also note that the Mach region 
is more complex than predicted in \cite{Satarov:2005mv}. Mach surface is not smooth. Inhomogeneity of the medium plays significant role is distorting the Mach region.  

Using the standard Cooper-Frey prescription, we now calculate $p_T$ integrated pion spectra $\frac{dN}{d\phi}$, 
at the freeze-out 

\begin{figure}[h] 
\includegraphics[bb=14 13 581 829,width=0.80\linewidth,clip]{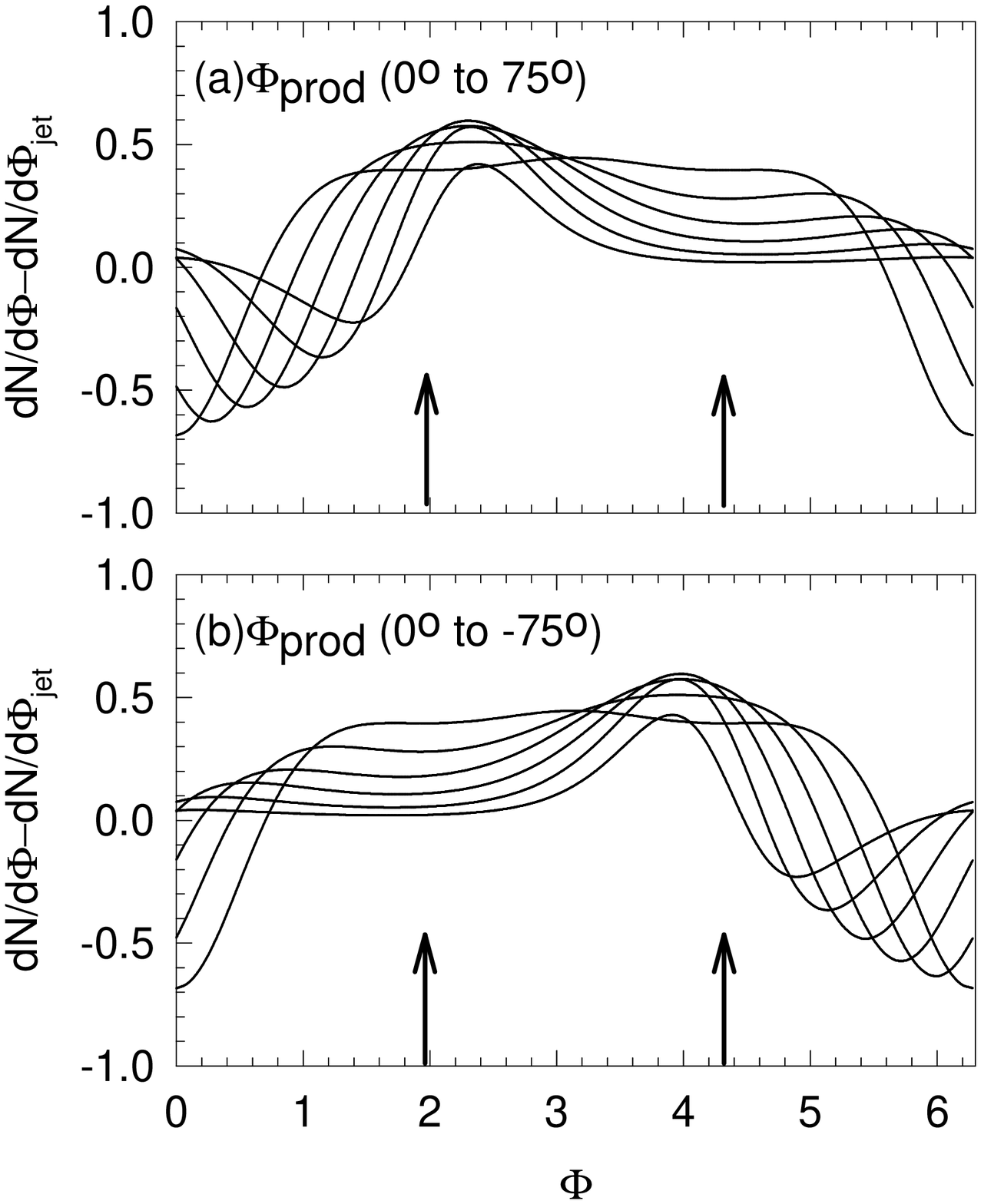}
\vspace{-3.3cm}
\caption{Azimuthal distribution of   pions due to quenching jet alone. In the upper panel, the quenching jet is confined to upper
half plane, $\phi_{prod}$=$0^o$ to $75^o$,
in steps of $15^o$ (from top to bottom). In the lower panel
the jet is confined to lower half plane,
$\phi_{prod}$=  $-75^o$ to $0^o$in steps of $15^o$.
The arrows indicate the Mach peaks ($\pi \pm \theta_M$) for resonance hadron gas ($c^2_s \approx 0.15$).}\label{F3}
\end{figure}

\noindent
surface at freeze-out
 temperature of 120 MeV.  
In Fig.\ref{F3}a,  for various position of the jet (confined to upper half plane) angular distribution of the pions is shown.
To show clearly the effect of deposited energy, we have subtracted the distribution obtained  in a evolution without a quenching jet.
 When the jet
is at the periphery of the upper half plane, $\phi_{prod}=75^o$, the angular distribution is hardly changed in the lower half plane.  Particle production is enhanced at $\phi\approx 0.85 rad$. We also observe depleted
 production at $\phi=\phi_{prod}$. The distribution donot show
any second peak.
As the jet trajectory comes closer to the centre, qualitatively, the distribution remains unchanged. The peak position remains nearly unaltered but it gets widened significantly. The depletion
of production around $\phi=\phi_{prod}$ continues. We also find
that as the trajectories comes closer and closer to the diagonal,
particle distributions in the lower half plane is also increased. However, the second peak eludes us. For 
the trajectories closer to the diagonal, the single peak observed
in off diagonal jet trajectories, gives way to a broad maxima.
A mirror image is found for the jets in the lower half plane,
 Fig.\ref{F3}b. 
The jet enhances the particle production around $\phi\approx 4rad$.
As before, the peak gives way to a broad maxima, as the jet 
trajectory comes closer and closer to the diagonal. Here also, we find depleted production at $\phi=\phi_{prod}$. Apparently, as the jet moves
in, it drags fluid along with it, resulting a reduced particle production around $\phi=\phi_{prod}$.

In the Fig.\ref{F3}, the arrows indicate the Mack peak positions,
$\phi=\pi\pm \theta_M$, for a static resonance hadron gas with
squared speed of sound $c^2_s\approx 0.15$.  
If we associate the excess production with shock wave
formation due to quenching jet \cite{chaudhuri}, interesting conclusions can be drawn. For a jet confined to upper half plane, the shock wave 
is distorted to such an extent that $\tilde{\theta}_{M-}\approx 0$ and $\tilde{\theta}_{M+}\approx 0.72rad$, while for the jets
confined to lower half plane, $\tilde{\theta}_{M+}\approx 0$ and
$\tilde{\theta}_{M-}\approx 0.72rad$. The angles should be compared with the cone opening angle $\tilde{\theta}_{M\pm}\approx 0.4rad$ for resonance hadron gas. Due to finite fluid velocity and
inhomgeneity of the medium, the inside shock front disappear,
while the outside shock front opened up significantly.
\begin{figure}[h] 
\includegraphics[bb=14 13 581 829,width=0.80\linewidth,clip]{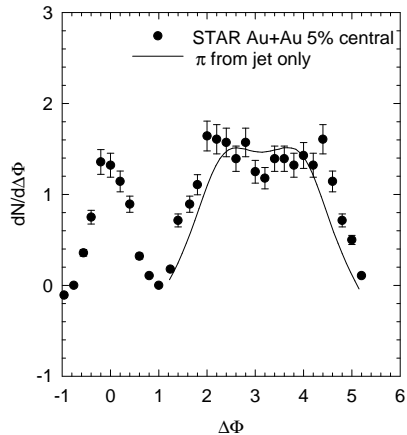}
\vspace{-4.3cm}
\caption{Black circles are STAR data on background subtracted secondaries in
azimuthal angle $\Delta\phi$, associated with hard trigger ($p_T > $ 4 GeV), in 0-5\% centrality Au+Au collisions. The solid line is the {\em normalised}
azimuthal distribution of excess pions, due to quenching jet only, in $b=3$ fm Au+Au collisions.} \label{F4}
\end{figure}

As mentioned in the beginning, STAR  collaboration, in 5\% central Au+Au collisions, has seen structures in the azimuthal distribution of background subtracted secondaries, associated with high $p_T$ ($p_T>$  4 GeV) trigger \cite{Wang:2004kf}.  
Are such structures associated with conical flow as suggested
in \cite{shuryak}? In Fig.\ref{F4} we have compared the 
preliminary STAR data \cite{Wang:2004kf} with (normalised) 
azimuthal distribution of pions due to quenching jet only.
The spectrum is obtained after averaging over all the possible jet trajectories.  The away side azimuthal distribution of back ground subtracted STAR data  closely match with the angular distribution of pions
due to quenching jet only.  We also reproduce the hint of depression at $\phi=\pi$.

To summarise, we have studied the Mach shock wave formation in central Au+Au collisions  due to a quenching jet.
Hydrodynamical equations, with a source (representing the quenching jet) are numerically solved with initial conditions appropriate for Au+Au collisions at b=3 fm and with jet energy loss,
consistent with observed high $p_T$ suppression. Explicit simulations
indicate that  the Mach shock front generated by the quenching jet, depend significantly on the jet trajectory. Finite fluid velocity
(and also inhomogeneity of the medium) cause significant distortions of the shock region. For a jet moving off-diagonally,   Mach cone become asymmetric with respect to jet axis. The inside shock front is pushed in, while the outside shock front is
pushed out.  Azimuthal distribution of pions, due to quenching jet only, suggest that for a off-diagonal jet trajectory, the distortions can be large, such that the  inside Mach shock front disappear.
We have also shown that the ({\em normalised}) azimuthal distribution of pions due to quenching jets only, averaged over all the trajectories, reasonably well explains the STAR data \cite{Wang:2004kf} on the background subtracted secondaries, associated with high $p_T$ trigger, along with the hint of depression at $\phi=\pi$. It strongly suggest that the structure is closely linked with "conical" flow due
to partonic energy loss.



\begin{thebibliography}{99}
\bibitem{chaudhuri}We have verified that the peaks in Fig.\ref{F3}, are indeed related to speed of sound of the medium. For a off-diagonal jet, peak position changes if the speed of sound of the medium is changed.
\end{thebibliography}
\end{document}